\documentstyle[epsfig,amsmath,amssymb,float]{elsart}

\begin{document}
\begin{frontmatter}

%%%%%%%%%%%%%%%%%%%%
\newlength{\defbaselineskip}
\setlength{\defbaselineskip}{\baselineskip}
\newcommand{\interlinea}[1]{\setlength{\baselineskip}{#1 \defbaselineskip}}

\newcommand\ket[1]{\left|#1\right\rangle}
\newcommand\bra[1]{\left\langle#1\right|}
\newcommand\lavg{\left\langle}
\newcommand\ravg{\right\rangle}
\newcommand\be{\begin{equation}}
\newcommand\ee{\end{equation}}
\newcommand\la{\leftarrow}
\newcommand\ra{\rightarrow}
\newcommand\mf{\mathbf}
\newcommand\trm{\textrm}
\newcommand\id{{\rm 1} 
\hspace{-1.1mm} {\rm I}
\hspace{0.5mm}}
\newcommand\bea{\begin{eqnarray}}
\newcommand\eea{\end{eqnarray}}
\newcommand\stocavg[1]{{\lavg #1 \ravg}_{\trm{stoc}}}
\newcommand\fk[1]{\mathfrak{#1}}

%%%%%%%%%%%%%%%%%%%%%%%%%%%%%%

\title{Slave boson model for two-dimensional trapped Bose-Einstein condensate}

\author[sns]{K. K. Rajagopal}
\corauth[cor1]{Corresponding author, e-mail: {\tt k.rajagopal@sns.it}}
\address[sns]
{Scuola Normale Superiore, I-56126 Pisa, Italy}
\maketitle
\begin{abstract} A system of N bosons in a two-dimensional harmonic trap is 
considered. The system is treated in term of the slave boson representation 
for hard-core bosons which is valid in the arbitrary density regimes. I 
discuss the consequences of higher order interactions on the density profiles 
by mapping the slave boson equation to the known Kohn-Sham type equation 
within the Density functional scheme.
\end{abstract}

\end{frontmatter}

\section{Introduction}

The possibility of Bose-Einstein condensation (BEC) phase transition occuring 
in a 2D geometry has a remarkable history and breakthroughs. Recent 
experiments have realized a quasi-two-dimensional (Q2D) cold Bose gas by 
tuning the anisotropy of the trapping potential \cite{gorlitz,safanov}. This 
achievement has attracted vast attention on the new physics appearing in the 
currently studied low-dimensional trapped gases and the nature of finite-size 
and finite-temperature effects. The effect of dimensionality on the existence 
and character of BEC or superfluid phase transition has seen a spurt of 
interest in recent years. A peculiar feature of the low-dimensionality (1D or 
2D geometry) at finite temperature is the absence of a 'true condensate' 
following the Bogoliubov $k^{-2}$ theory, believed to be originated from the 
long wave-length phase fluctuations \cite{popov}. This has 
been proved in the seminal work of Mermin and Wagner \cite{mermin} and 
Hohenberg \cite{hohenberg67}. A Kosterlitz-Thouless transition to a superfluid 
state still occurs on cooling through the binding of vortices of opposite 
vorticity \cite{kosterlitz} and leads to an algebraic decay of the one-body 
density matrix. However, in a trapped 2D fluid the modification of the density 
of states caused by the confining potential allow a true condensate to exist 
even at finite temperature. Another important consequence lower dimensionality
is that the T-matrix  for two-body collisions {\it in vacuo} at low momenta 
and energy in the dilute regime vanishes \cite{popov,schick}. It is then 
necessary to evaluate the scattering processes between pairs of Bose particles 
by taking into account the presence of surrounding condensate particles 
\cite{stoof93,bijlsma97,Lee,rajagopal}. \par

The dilute limit ($\rho a^2\ll\,1$), where $\rho$ is the uniform density of 
particles and $a$ is the s-wave scattering length of a Bose gas can be 
described quite satisfactorily by traditional approaches, such as the 
Gross-Pitaevskii(GP) approach in the weakly interacting regime \cite{gross61}. 
However, it is intuitively obvious that as the density of particles in the 
condensate or the scattering length increases, the nature of interactions will 
become more and more complex, and one will need to take even three-body and 
higher order effects into account. It is a crucial point since in the strictly 
two-dimensional trap, the scattering length of the bosons become larger than 
the axial thickness of the condensate cloud. Thus we need to go beyond the 
mean-field approach to describe the system of dense Bose gas .\par

These limitations motivates me to address the problem particularly the 2D 
dense condensate by introducing the slave boson model (SB) which is valid for 
arbitrary density regimes. The slave boson technique was first 
introduced by Kotliar and Ruckenstein \cite{kotliar}, who used it to deal with 
the fermionic Hubbard model. A functional integral approach to the problem of 
hard-core bosons hopping on a lattice has been previously put forward by 
Ziegler \cite{ziegler93} and Fr{\'e}sard \cite{fresard}. Ziegler 
\cite{ziegler93} has demonstrated that slave boson (SB) representation allows 
one to describe the dynamics of Bose gas taking into account three-body and 
higher effects at arbitrary densities for the three-dimensional (3D) case. The 
advantage of using slave boson representation is by the fact that there are 
only two states per site in a hard-core system: a lattice site is either empty 
or occupied by a single boson. A hopping process of a boson appears as an 
exchange of an empty site with singly occupied site.\par

The paper is outlined as follows. In section 2 the Bose gas with hard-core 
interaction defined as an effective functional integral form is summarized
in the slave boson formalism. The SB equation is evaluated in section 3. 
Some concluding remarks are presented in section 4. \par

\section{The Slave Boson model}

A Bose gas, defined as a grand canonical ensemble of bosons, can be described 
in a second quantization formalism, for instance using a functional integral 
representation \cite{negele} as following:
\begin{equation}
Z=\int\exp(-S[\phi^{*},\phi]/\hbar)\prod d[\phi^{*}]d[\phi] 
\label{funcInt}
\end{equation}
where the Euclidean action $S[\phi^{*},\phi]$ of the complex field $\phi$ 
needs to be defined accordingly to the model system of interest ({\it e.g.} 
GP, SB, ect.). In this work I will resort to the slave boson representation of 
hard-core bosons which has been exhaustively defined in the literatures 
\cite{kotliar,ziegler93,ziegler97,dicke}. The representation relies on the 
Bose-Hubbard model where the particle trades its position with an empty site 
on the lattice. Both particle as well as the empty site are described by the 
corresponding creation and annihilation operators. In the slave boson
representation for instance, a lattice site $x$ that is occupied by boson is 
represented by complex field $b_{x}$ and a site that is empty by $e_{x}$. Thus 
the slave boson action $S_{sb}$ can be translated into the following picture 
\cite{kotliar,barnes}
\begin{equation}
S_{sb}=\sum_{x,x'}b_{x}^{*}e_{x}t_{x,x'}b_{x'}e_{x'}^{*}+V_{x}b_{x}^{*}b_{x}
+i\lambda_{x}(e_{x}^{*}e_{x}+b_{x}^{*}b_{x}-1)\,
\label{sbaction} 
\end{equation} 
where the first term describes the exchange of bosons and empty sites in a 
hopping process at sites $x$ and $x'$ with rate $t_{x,x'}$. In the second term 
I have introduced an external confining potential $V_{x}$. The field 
$\lambda_{x}$ in the last term of the above equation  enforces a constraint 
$e_{x}^{*}e_{x}+b_{x}^{*}b_{x}=1$ which guarantees that a site is 
either empty or singly occupied. It is indeed the core condition of the slave 
boson model. With this choice of action, the functional integral in 
Eq. (\ref{funcInt}) can be written explicitly describing the partition 
function of the grand canonical ensembles of bosons as:
\begin{equation}           
Z=\int e^{-S_{sb}}\prod_{x}d\lambda_{x}db_{x}db_{x}^{*}
de_{x}de_{x}^{*}.
\label{sbInt}
\end{equation}
The slave boson field $e_{x}$ and $b_{x}$ can be combined to form a 
collective field $b_{x}^{*}e_{x}\rightarrow \Phi_{x}$. Then the 
constraint field $\lambda_{x}$ and the slave bosons fields can be 
integrated out which finally leads to the action for the collective field 
$\Phi_{x}$. This was demonstrated in detail in \cite{ziegler93,ziegler97}.
Here, I only present the results: 
\begin{equation}
Z=\int e^{-S_{b}-S_{1}}\prod_{x}  d\Phi_{x}d\Phi^{*}_{x} 
\label{sbInt2}
\end{equation}
with  the hopping("kinetic") term
\begin{equation} 
S_{b}=\sum_{ x, x'}\Phi_{x}(1-t)^{-1}_{x,x'}\Phi^{*}_{x'}
\approx\sum_{x}(1-\tau)^{-1}|\Phi_{x}|^{2}+\frac{\tau}{6a^{2}
(1-\tau)^{2}}\Phi_{x}(\nabla^{2}\Phi^{*}_{x})
\label{sbhopp}
\end{equation}
and the potential term 
\begin{equation}
  S_{1}=-\sum_{x}\log \left( e^{-V_{x}-1/4}\int 
d\varphi \exp[-\varphi^2]\frac{\sinh\sqrt{(\varphi-V_{x}^2)^{2}
+|\Phi_{x}|^2}}{\sqrt{(\varphi - V_{x}^2)^{2}+|\Phi_{x}|^2}}
 \right ).
\label{sbpot}
\end{equation}
$V_{x}= x^2$ is the planar harmonic confinement where motion of atoms are in 
planar direction with dimensionless radial distance $x$. The density of the 
condensate in the trap can be calculated by the saddle-point approximation. In 
this work, we neglect the thermal fluctuations which is not important at zero 
temperatures for Bose-Einstein condensation to occurs in the system. However,
if one needed to account for the thermal fluctuations, one could do so by 
studying the deviations around the saddle point in the functional integral of 
Eq. (\ref{sbInt2}) \cite{negele}. That means, we have to look for the solution 
order parameter $\Phi_{x}$ which minimizes $S_{b}+S_{1}$ taking a 
saddle-point approximation on the functional integral Eq. (\ref{sbInt2}) to 
obtain the generalized non-linear Schr\"odinger equation \cite{ziegler97} 
\begin{equation}
\left( \frac{\tau}{6a^2}\nabla^2+ \alpha + \alpha^2\frac
{\partial S_{1}}{\partial |\Phi_{x}|^2}\right)\Phi_{x}=0 \,,
\label{SB}
\end{equation}
where $\alpha = 1-\tau$. The term $\partial S_{1}/\partial 
|\Phi_{x}|^2$ contains higher-order interaction (three-body 
interactions, {\it etc.}). However this equation is difficult to solve as it is
because the term $\partial S_{1}/\partial |\Phi_{x}|^2$ can lead to heavy 
non-linearity.

\section{Density functional solution to the slave boson equation}

\begin{figure}
\centering{
\epsfig{file=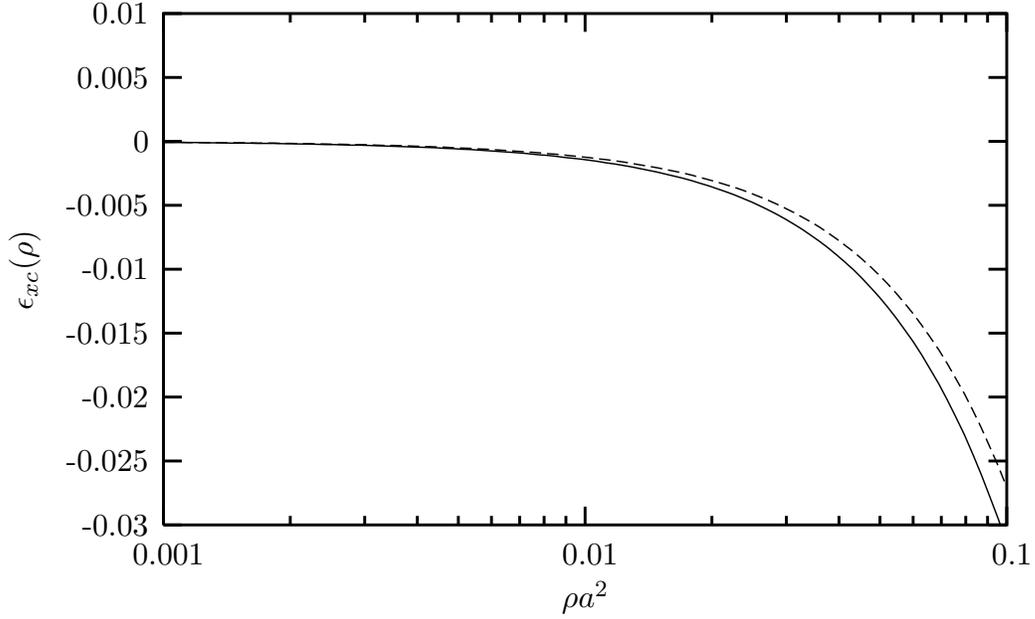,width=1.0\linewidth}}
\caption{Homogeneous exchange correlation energy $\epsilon_{xc}[\rho]$ versus 
the gas parameter $\rho a^{2}$ for the expression $\epsilon_{xc}^{kol}$ 
(solid line) compared to $\epsilon_{xc}^{vmc}$ (dashed lines)}
\label{figcoup}
\end{figure}

Taking into account the possible effects of higher order interactions and to 
avert strong non-linearity in Eq. (\ref{SB}), I find a much feasible
way to solve it by mapping to the known Kohn-Sham \cite{KS} type equation 
$via$ Density functional theory (DFT) scheme. The main idea of the DFT 
formalism is in finding an equilibrium ground state energy by minimizing the 
following energy functional
\begin{equation}
 E[n(x)]=  T[n(x)] + E_{ex}[n(x)]
 +\int d{x}\, n(x)V_{x}
\label{energy}
\end{equation}
which is a unique functional of density $n(x)=|\Phi_{x}|^2$ for a given 
external potential $V_{x}$ \cite{HK64}. The first term in Eq. (\ref{energy}) 
is the kinetic energy of the non-interacting system while the second term 
represent the excess contribution of free energy $E_{ex}[n(x)]$. The 
remainder term is the interaction energy of the particles in an external 
field. Minimizing $E[n(x)]$, the ground state energy of the system can 
be determined by imposing the stationary equlibrium condition
\begin{equation}
\frac{\delta E[n(x)]}{\delta n(x)}=\mu
\label{vari}
\end{equation} 
where the Lagrange multiplier $\mu$ can be identified as the chemical 
potential. This is equivalent to the condition of equilibrium for a 
non-interacting particles under the influence of an  
effective external potential \cite{moroni}
\begin{equation}
v_{eff}[n(x)]=\frac{\delta E_{ex}[n(x)]}{\delta n(x)}+V_{x}\,.
\label{eff}
\end{equation}
It is evident that the equilibrium density profile $n(x)$ can be obtained
for a given $v_{eff}[n(x)]$ by solving the following Kohn-Sham type 
equation \cite{KS},
\begin{equation}
\left[\frac{\delta T[n(x)]}{\delta n(x)}+v_{eff}[n(x)]\right]
\Phi_{x}=\mu\Phi_{x}\,.
\label{KS}
\end{equation} 
By setting $\alpha=-\mu$ and $\tau/6a^{2}=-1$, Eq.(\ref{SB}) can be mapped to 
Eq. (\ref{KS}) yielding ,
\begin{equation}
 \mu^2 S_{1}[n(x)]= \hat{V}_{H}[n(x)]+E_{xc}[n(x)]
 + \int d{x}\,n(x)V_{x}\,.
\label{relation}
\end{equation}
In this relation, I have introduced an exchange-correlation energy 
$ E_{xc}[n(x)]$ by explicitly splitting from the excess free energy 
$ E_{ex}[n(x)]$  the following Hartree energy \cite{moroni},
\begin{equation}
\hat{V}_{H}[n(x)]=\frac{1}{2} \int \int \,d{x}\,d{x'} 
V({x}-{x'}) n(x)n(x')\,
\label{Hartree}
\end{equation}
with a generic inter-particle potential $V({x}-{x'})$. In general, the 
exchange correlation functional energy  $E_{ex}[n(x)]$ in a DFT calculation is 
not known exactly. Despite, one can resort to approximations such as the Local 
density approximation (LDA) which reads,
\begin{equation}
 E_{xc}[n(x)]\approx \int\,d{x} E_{xc}^{hom}[\rho]|_{\rho\rightarrow n(x)}\,
\label{LDA}
\end{equation}
where $E_{xc}^{hom}[\rho]=\rho\epsilon_{xc}[\rho]$  and $\epsilon_{xc}[\rho]$ 
is the exchange correlation energy of a homogeneous system with uniform 
density $\rho$. The Functional derivatives of the above relation can be 
written as \cite{albus}
\begin{equation}
V_{xc}[n(x)]=\frac{\delta E_{xc}[n(x)]}{\delta  n(x)}=
\frac{\partial (\rho \epsilon_{xc}[\rho])}{\partial \rho}|_{\rho\rightarrow 
n(x)}\,.
\label{partial}
\end{equation}
Hence Eq. (\ref{SB}) can be written explicitly as the following Kohn-Sham 
type equation:
\begin{equation}
\left[-\nabla^2 + r^2 + V_{H}[n(x)]+V_{xc}[n(x)]\right]
\Phi_{x}= \mu\Phi_{x}\,,
\label{KS2}
\end{equation}
where $V_{H}[n(x)]= 4n(x)/|\ln(n(x)a^2)|$ is the Hartree potential 
calculated from Eq. (\ref{Hartree}) using a hard-core potential 
\cite{schick,tanatar}. The homogeneous exchange correlation energy 
$\epsilon_{xc}[\rho]$ can be approximated to the following expression
\cite{kolomeisky}: 
 \begin{equation}
 \epsilon_{xc}^{kolo}[\rho]\approx 
\frac{2\rho}{\ln(1/\rho a^2)}\left (-\frac{\ln(\ln(1/\rho a^2))}
{\ln(1/\rho a^2)}\right )\,.
\label{kolo}
\end{equation}  
Very recently Pilati et. al \cite{pilati} did a variational Monte Carlo (vmc) 
calculation in obtaining the ground state energy of 2D bose gas using various 
inter-atomic potentials. The following is the fit of their beyond mean-field 
ground state energy : 
\begin{equation}
 \epsilon_{xc}^{vmc}[\rho] \approx 
\frac{2\rho}{[\ln(1/\rho a^2)]}\left (-\frac{0.86\ln(\ln(1/\rho a^2))}
{\ln(1/ \rho a^2)}\right )\,.
\label{vmc}
\end{equation}  
Fig (\ref{figcoup}) illustrate these homogeneous exchange correlation energies 
for various values of gas parameter $\rho a^{2}$. Setting $E_{xc}[n(x)]=0$ one 
recovers the GP equation.\par 

There are three cases to consider in solving  Eq.(\ref{KS2}), namely using 
information $\epsilon_{xc}^{kolo}[\rho]$, $\epsilon_{vmc}^{kolo}[\rho]$ or 
$E_{xc}[n(x)]=0$. The KS equation Eq.(\ref{KS2}) need to be solved 
self-consistently with the condition that the areal integral of the condensate 
density $n(x)=|\Phi_{x}|^2$ is equal to the total number $N$ of particles. For 
a numerical illustration, we have taken the radial trap frequency, and the 
scattering length as appropriate for $^{23}$Na atoms in the experiment of 
G\"orlitz {\it et al.} \cite{gorlitz}, namely $\omega_{\perp}=188.4\,$Hz, 
s-wave scattering length $\tilde{a}= 2.8$ nm. I am implicitly assuming 
however, that the trap has been axially squeezed to reach the strictly 2D 
scattering regime. In this work all the distances and energies 
have been scaled in the harmonic oscillator unit 
($a_{ho}=\sqrt{\hbar/m\omega_{\perp}}$) and energies by 
$\hbar\omega_{\perp}/2$. Numerical iteratation method by discretization using 
a two-step Crank-Nicholson has been used in performing this calculation for 
$5\times 10^{3}$  and $5\times 10^{4}$ $^{23}$ sodium atoms.

\begin{figure}
\centering{
\epsfig{file=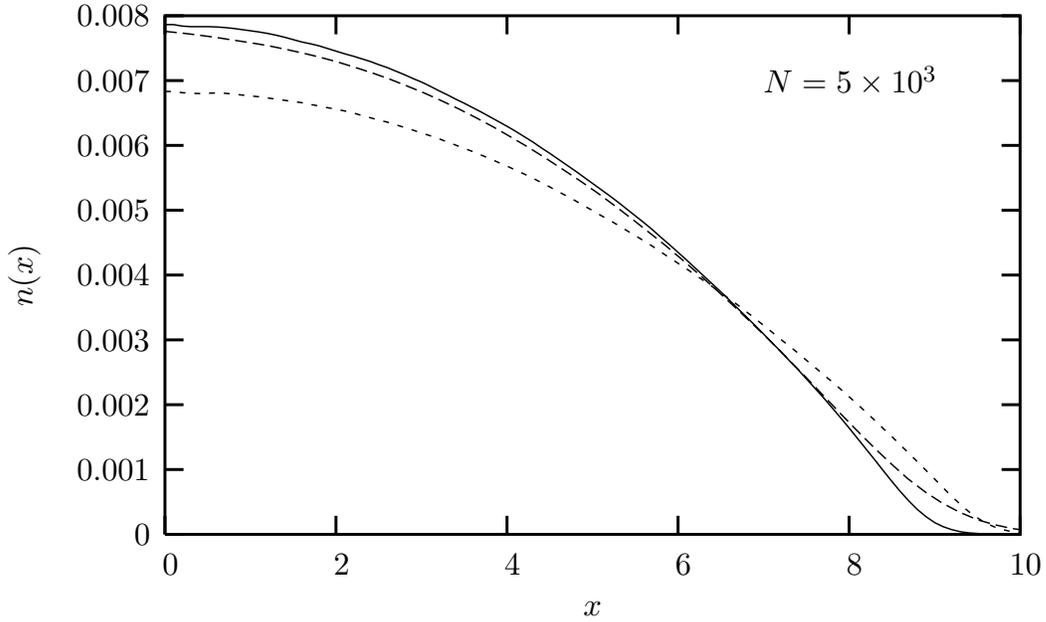,width=1.0\linewidth}}
\caption{Density profiles $n(x)$ for the condensate (in units $a_{ho}^{-2}N^{-1}$, with $a_{ho}=\sqrt{\hbar/m\omega_{\perp}}$) versus the radial distance 
$x$ (in units of $a_{ho}$) obtained by solving Eq.(\ref{KS2}) using information
of $\epsilon_{xc}^{kol}$ (solid line), $\epsilon_{xc}^{vmc}$ 
(long-dashed lines) and setting $E_{xc}[n(x)]=0$ (short-dashed lines).}
\label{fig5P3}
\end{figure}

\begin{figure}
\centering{
\epsfig{file=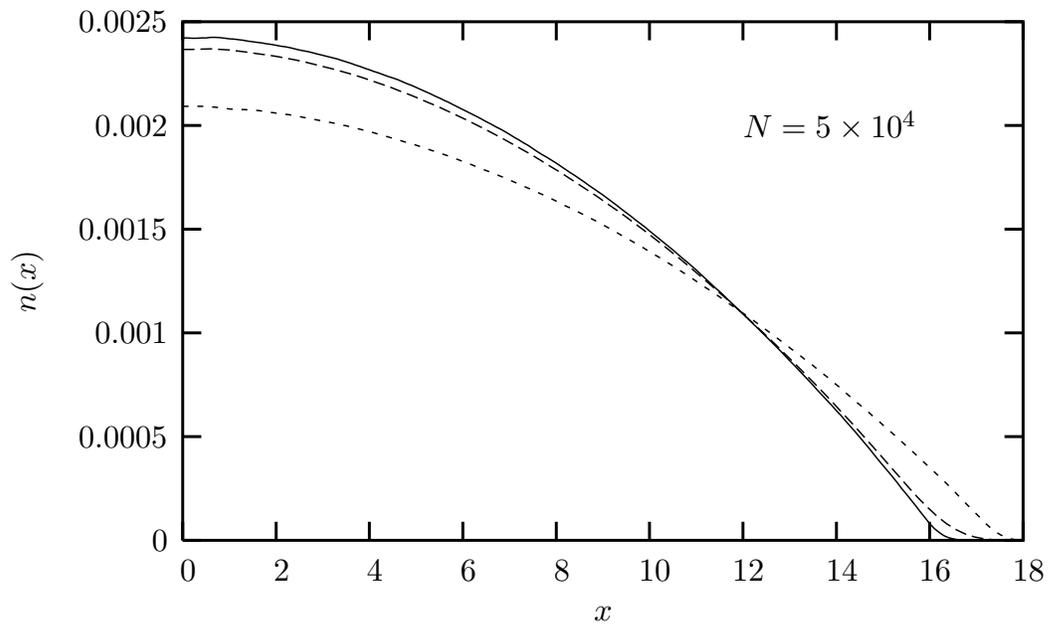,width=1.0\linewidth}}
\caption{caption as in Fig. (\ref{fig5P3}) for $N=5\times 10^{4}$ atoms.}
\label{fig5P4}
\end{figure}

The results obtained from Eq. (\ref{KS2}) using information of the exchange 
correlation energies $\epsilon_{xc}^{kol}[\rho]$ [$n_{kol}(x)$] and  
$\epsilon_{xc}^{vmc}[\rho]$ [$n_{vmc}(x)$] comparing to the case with 
$E_{ex}[n(x)]=0$ [$n_{GP}(x)$] are depicted in Fig. (\ref{fig5P3}) and 
Fig. (\ref{fig5P4}) for  $5\times 10^{3}$ and $5\times 10^{4}$ atoms 
respectively. The main effect of the correlation potentials can be observed 
where the condensate density reaches the peak at the trap center. So it is 
feasible to explain the observation by measuring the differences 
$\Delta n^{*}=n_{kol}(0)-n_{GP}(0)$ and 
$\Delta \tilde{n}=n_{kol}(0)-n_{vmc}(0)$ at the trap centre as done 
by Nunes \cite{nunes}. For a fixed number of atoms $N$ the conditions 
$\int d{x} \Delta n^{*}=0$ and $\int d{x} \Delta \tilde{n}=0$ are imposed so 
that the rest of curves $\Delta n^{*},\Delta \tilde{n}$ are well determined 
at $x=0$. The differences $\Delta n^{*},\Delta \tilde{n}$ increases with
$\Delta n^{*}/n_{kol},\Delta \tilde{n}/n_{kol}$ varying from 9$\%$ and 
1.33$\%$ for $N=5\times 10^{3}$ reaching about 13$\%$ and 2.27$\%$ 
for $N=5\times 10^{4}$ respectively. The increment in 
$\Delta n^{*},\Delta \tilde{n}$ corresponds to the magnitude of reduction in 
the effective interaction energy of the system. The small increment in 
$\Delta \tilde{n}$ shows that the density profile predicted using the 
correlation energies $\epsilon_{xc}^{vmc}$  agrees well with the one obtained
using $\epsilon_{xc}^{kol}$. However what is more spectacular is 
the large values of $\Delta n^{*}$ and the increment of 
$\Delta \tilde{n}/n_{kol}$ from 9$\%$ to 13$\%$ as the number the number of 
atom increases from $N=5\times 10^{3}$ to $N=5\times 10^{4}$. The effect of 
exchange-correlation clearly shows a reduction in effective interaction energy 
thus the profiles are more localized at the centre of the trap in comparison 
to the GP case. This argument can also be supported by the observation in 
Fig (\ref{figcoup}) where the magnitude of homogeneous exchange correlation 
energies decreases as the density increases. On the other hand 
$\epsilon_{xc}[\rho]$ is negligible for the very dilute regimes 
$\rho a^{2} \ll 0.01$ justifying GP approximation is only a good model in this 
limit

\section{Conclusion}

In summary, the harmonically trapped 2D Bose gas at zero temperature 
described by the slave boson model has been studied for both the dilute and 
dense limits. The central issue of this work is in how to treat the slave 
boson term $\partial S_{1}/\partial |\Phi_{x}|^2$ in the highly non-linear
partial differential equation (\ref{SB}). The importance of the higher order 
interactions $via$ the exchange-correlation energy has been revealed by 
mapping Eq. (\ref{SB}) to the Kohn-Sham scheme of DFT in the dense limit. This 
latter term is responsible for the correction of up to 9$\%$ and 13$\%$ of the 
peak density for $N=5\times10^{3}$ and $N=5\times10^{4}$ atoms at the center
of trap. In other words these exchange correlation has restored the correct 
high density behaviour of the inhomogeneous Bose condensate imposed by the 
strict hard-core condition of the slave-boson representation. On the opposite
dilute regime I have demonstrated that the slave boson equation (\ref{SB}) can 
be approximated to a Gross-Pitaevskii type equation ({\it via} DFT).

At absolute zero temperature where the thermal fluctuation is frozen out and 
quantum fluctuation is prominent. This quantum fluctuation plays an important 
role in the phase transition from superfluid to mott insulator 
\cite{sachdev,greiner,zwerger} in the three-dimensional case. Recently Wu 
et.al. \cite{wu} have demonstrated that a vortex in 2D optical lattice can
induce the SF-MI transition. As a future direction a practical realization of 
this transition will be investigated using the slave-boson model 
of the 2D condensate accomodating a single or many vortices.

\subsection*{Acknowledgement}

I would like to thanks  Professor Ziegler, Professor S. K. Adhikari and
Ch. Moseley for useful comments. Also acknowledging Inst. f{\"u}r Physik, 
Augsburg for the hospitability and financial support during early part of this 
work.

\clearpage

\clearpage

%\bibliography{slave}
%\end{document}

\end{document}